\documentclass[nofootinbib,apsspec,twocolumn]{revtex4-1}
\usepackage{amsmath}
\usepackage{amssymb}
\usepackage{graphics}
\usepackage[dvips]{graphicx}
\usepackage{graphicx}
\usepackage[usenames]{color}

\usepackage[colorlinks=true,linktocpage=true,linkcolor=blue,citecolor=blue]{hyperref}
\usepackage{hyperref}

\newcommand \beq{\begin{eqnarray}}
\newcommand \eeq{\end{eqnarray}}
\def\simge{\mathrel{%
       \rlap{\raise 0.511ex \hbox{$>$}}{\lower 0.511ex \hbox{$\sim$}}}}
\def\simle{\mathrel{
       \rlap{\raise 0.511ex \hbox{$<$}}{\lower 0.511ex \hbox{$\sim$}}}}

\begin{document}
\title{The Golden Era of Neutron Stars: from Hadrons to Quarks\footnote{Proceedings, 8th Intl. Conference on Quarks and Nuclear Physics 2018 (Tsukuba, Japan).
 }}
\author{Gordon Baym} 
\affiliation{\mbox{Department of Physics, University of Illinois, 1110
  W. Green Street, Urbana, IL 61801, U.S.A.} \\
}

\begin{abstract}

  Neutron stars were first posited in the early thirties, and discovered as pulsars in late sixties; however, only recently are we beginning to understand the matter they contain.  This talk describes the continuing development of a consistent picture of the liquid interiors of neutron stars, driven by four advances:  observations of two heavy neutron stars with masses $\simeq$ 2.0 solar masses; inferences of masses and radii simultaneously for an increasing number of neutron stars in low mass X-ray binaries, and future determinations via the NICER observatory; the observation of the binary neutron star merger, GW170817, through gravitational waves as well as across the electromagnetic spectrum; and an emerging understanding in QCD of how nuclear matter can turn into deconfined quark matter in the interior.  We describe the modern quark-hadron crossover equation of state, QHC18, and the corresponding neutron stars, which agree well with current observations.  
\end{abstract}

\maketitle

\section{Introduction}

     The recent past has seen remarkable progress in neutron-star physics, converging from many directions \cite{NSReview} and
providing new observational windows into the microscopic nuclear physics of dense matter.    First, two heavy neutron stars, with masses $\simeq$ 2.0 solar masses  ($M_\odot$), have been cleanly identified \cite{fonseca,Demorest,Antoniadis2013};  how dense matter can support such a large mass against gravitational collapse challenges our current understanding of the microscopic physics of matter, from baryon densities of several to 10 $n_0$, where $n_0\simeq$ 0.16 fm$^{-3}$ is the saturation density of nuclear matter.    Second, a program of simultaneous determination of masses and radii of neutron stars undergoing thermonuclear X-ray bursts or in quiescence  has been underway for a number of years \cite{OzelFreire,Ozel2010,Steiner2012,Ozel2015,Steiner2016,fredcole} (and references therein), further constraining the equation of state of dense matter.   Third, the NICER (Neutron Star Interior Composition Explorer) X-ray timing observatory, flown to the International Space Station in June 2018,
 will shortly begin to provide effectively model-independent precision data on masses and radii of neutron stars, which will enhance our knowledge of the equation of state of matter in neutron stars \cite{nicer}. The key to NICER is the bending of light by the gravitational field of the neutron star being observed,. which enables one to ``see" a bit behind the star and thus learn most simply its ratio of mass, $M$, to radius $R$.    And fourth, the spectacular detection by LIGO with Virgo of the merger of two neutron stars -- the event GW170817 -- through the gravitational radiation emitted in the merger \cite{GW170817,Abbott2018},  and seen across the entire electromagnetic spectrum from gamma rays through to radio 
 \cite{berger,EM170817} (and references therein),  has given us new ways of inferring both equilibrium and dynamical properties of neutron stars.

        On the theory side,  a better understanding in quantum chromodynamics (QCD) of how nuclear matter turns into deconfined quark matter is emerging, both at non-zero temperature at low density via lattice gauge theory \cite{WB,HotQCD}, and at high density, providing insights into the equation of state of matter in neutron star interiors \cite{kpsb}.  As the only  ``laboratory" in which one can study and test microscopic theories of cold dense matter in QCD, the combined observational and theoretical approach is complementary to probes of the states of hot dense matter in ultrarelativistic heavy ion collision experiments at RHIC in Brookhaven, the LHC at CERN, and at GSI in the future \cite{fair}.

\section{Neutron Star Observations}

\subsection{Heavy neutron stars}

  The discoveries of two massive neutron stars in binary orbits with white dwarf companions has important implications for dense matter in QCD.  The two neutron stars are the binary millisecond pulsar J1614-2230, with mass $1.928\pm0.017 M_\odot$ \cite{fonseca} (the original mass measurement was $1.97\pm  0.04 M_\odot$ \cite{Demorest}), and the pulsar J0348+0432 with mass $2.01 \pm 0.04 M_\odot$ \cite{Antoniadis2013}. Beyond these two stars, a number of extreme {\em black widow} pulsars (pulsars that are consuming their light companion stars through accretion)
may prove to have even higher mass, possibly as large as 2.5$M_ \odot$.  Candidates include  the  millisecond pulsars PSR J1957+20 \cite{vankerwijk}, PSR J2215+5135  \cite{Schroeder-Halperin}, and  PSR J1311-3430 \cite{Romani2012,Romani2015}; however the inference of their masses remains uncertain in the absence of more complete modeling of the heating of the companion stars by the neutron stars.   

\begin{figure}[h]
\vspace{-30pt}
\includegraphics[width = 8.5cm]{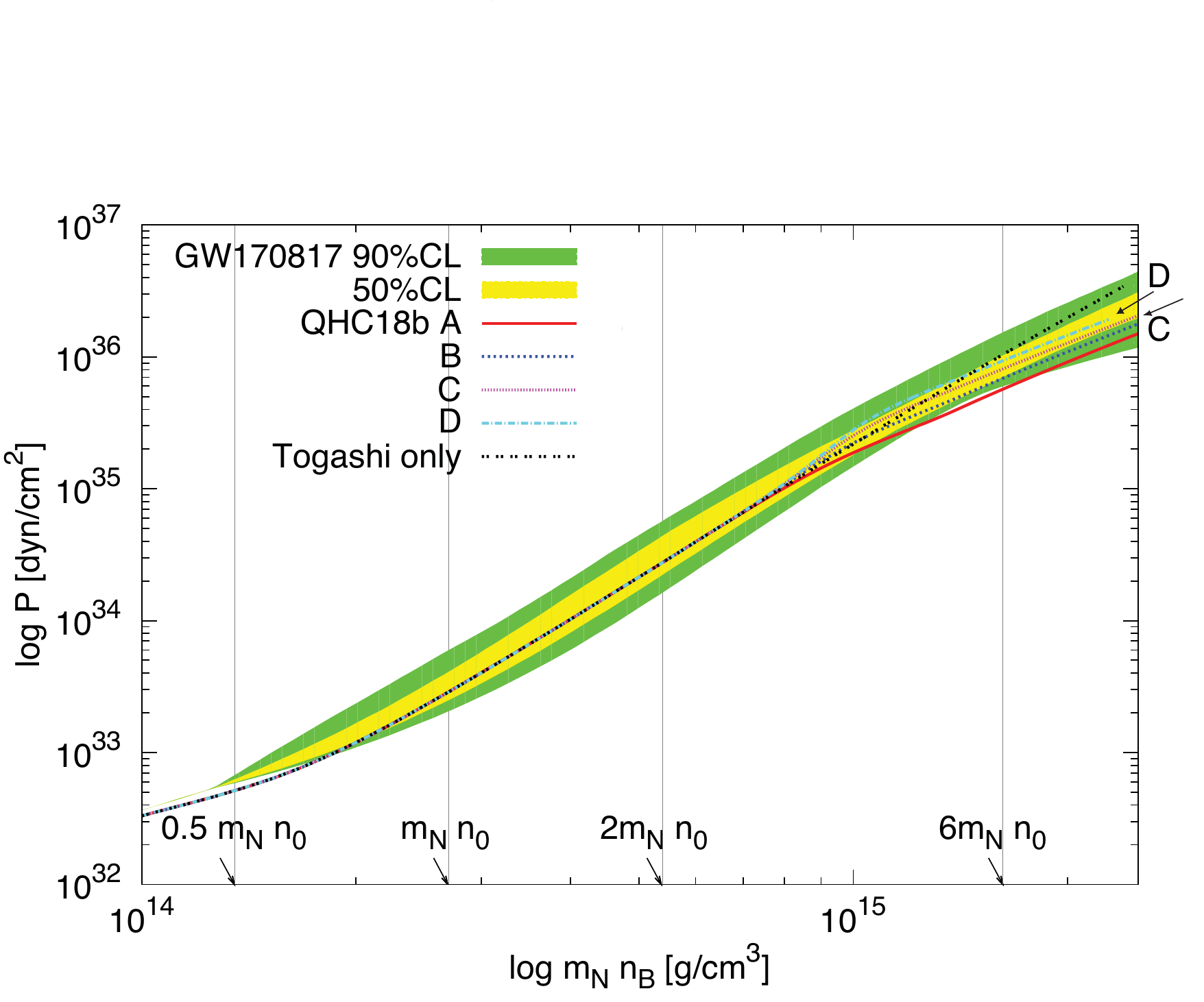}
\caption{\footnotesize{Comparison of the QHC18 equation of state for neutron stars, in pressure vs. baryon density, with the regions inferred by LIGO/Virgo \cite{Abbott2018}.   The letters A, B, C, and D correspond to choice of parameters for the short range universal quark repulsion $g_V$ and diquark pairing strength $H$ in quark matter, as shown in Fig. 5.  The parameter sets C through D agree remarkably well with the LIGO/Virgo inference.    Figure courtesy of T. Kojo ~\cite{hotqhc0}.}  } 
 \label{Ligo-QHC}
\end{figure}

      One immediate consequence of the existence of 2$M_\odot$ stars is that the neutron star equation of state must be stiff, with sufficiently large pressure for a given energy (or mass) density;  thus one can eliminate all equations of state that are too soft to explain these neutron stars.   
In addition, 2$M_\odot$ stars are particularly difficult to understand in terms of hadronic models of dense matter 
with hyperons added into the mix at about 2$n_0$, since strange hadrons generally soften the equation of state and limit the maximum stable star mass.    This ``hyperon puzzle" and possible resolutions are discussed in detail by Weise  \cite{weise}.
On the other hand, allowing strange quarks into quark matter, as mentioned below, presents no difficulties.

\subsection{Binary neutron star mergers}

      The GW170817 binary neutron star merger, seen in the galaxy NGC 4993 in the constellation Hydra in the Southern Hemisphere,
has broad implications for nuclear physics.   From the gravitational wave signal one has been able to infer a total mass of the two neutron stars of
2.73 (+0.04 -0.01) $M_\odot$, and individual stellar masses 1.36-1.60$M_\odot$, and 1.16-1.36$M_\odot$, 
infer limits on the tidal deformability of the neutron stars, as well as constraints on the equation of state of the matter they contain \cite{Abbott2018}; see Fig. 1.   And the subsequent electromagnetic radiation reveals the formation of elements heavier than Fe; indeed, the merger appears to be the primary site of the rapid neutron capture r-process in nucleosynthesis.   Just 1.7 sec. after the initial gravitational wave burst, marked by an inspiral and chirp, the Fermi Gamma-ray Burst Monitor (GBM) and, later, INTEGRAL saw gamma rays from the merger.   A series of independent observations of electromagnetic activity revealed the formation of a {\em kilonova} \cite{kilonova} in which the hot 2.7$M_\odot$ merged object produced, via the r-process, some 0.1 $M_\odot$ of elements heavier than Fe, including  Pt, Au, Th, U, and Pu -- as inferred by the radioactive decay of r-process elements dominating the UV, optical, and IR emission.  Binary neutron star mergers appear to be the predominant source of the gold in the universe,  with some 10-15 earth masses worth produced in this one event.   Particularly exciting is that many more binary neutron star mergers, perhaps up to 50 per year, will be discovered in the near future, through the combined efforts of LIGO, Virgo, and soon KAGRA.

\section{Inside Neutron Stars}

\subsection{The hadronic regime}

\begin{figure}[h]
\begin{center}
\includegraphics[width=9cm]{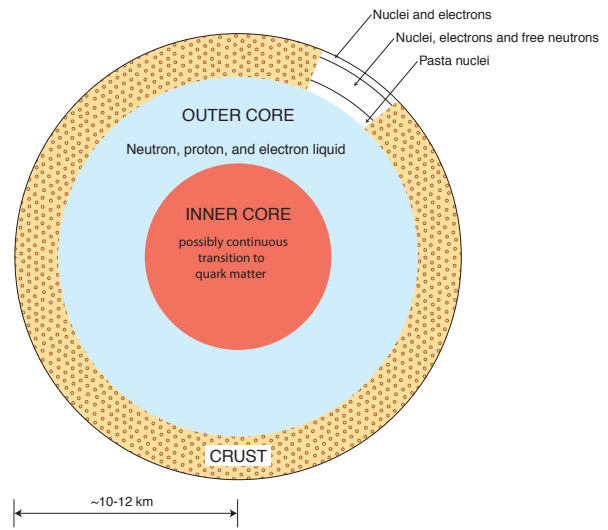}
\end{center}
\caption{\footnotesize{Schematic (not to scale) of a neutron star interior, showing the outer crust, the inner crust neutron drip regime with a sea of free neutrons, the region of non-spherical {\em pasta} nuclei in the inner crust; and in the outer core a hadronic liquid, and a quark liquid in the inner core.  } } 
 \label{nstar}
\end{figure}

     Let us turn to the interiors of neutron stars, focussing on open physics issues.  Measured masses of neutron stars are in the range $\sim$1.1-2.0$M_\odot$, corresponding to a baryon number $A \sim 1.2-2.4 \times10^{57}$.   Their general structure is shown in Fig.~\ref{nstar}, and reviewed in detail in Ref.~\cite{NSReview}.     

 With increasing density in the crust
 the matter becomes more and more neutron rich, changing from $^{56}$Fe (ideally) at the surface  to $^{62}$Ni  and then $^{64}$Ni.   Beyond this the standard nuclear shell model predicts a sequence of nuclei with a 50 neutron closed shell, ranging from $^{84}$Se to $^{76}$Ni, followed by a sequence of very neutron rich nuclei with an 82 neutron closed shell, from $^{124}$Mo  to $^{118}$Kr.  However, the magic numbers depend critically on spin-orbit forces, which are modified by three-body and tensor forces in nuclei, and are expected to be modified with increasing neutron richness.  Whether the magic numbers of such highly rich nuclei remain 50 and 82 is open, both experimentally and theoretically, see, e.g., \cite{otsuka} (and references therein).  Future experiment at rare isotope facilities, such as RIKEN, should be able to answer this question in the not too distant future.

     \begin{figure}[h]
\vspace{-116pt}
\begin{center}
\includegraphics[width=14cm]{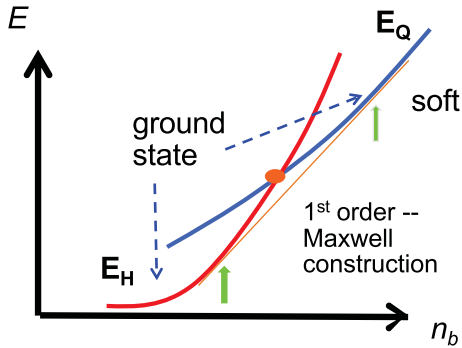}
\end{center}
\vspace{-60pt}
\caption{\footnotesize{Maxwell construction between hadronic and quark matter equations of state. The matter would make a jump from lower density hadronic matter to higher density quark matter, as indicated by the two green arrows. }} 
 \label{maxwell}
\end{figure}        
   
     At the neutron drip point the crust becomes, with depth, permeated by 
 a sea of free neutrons, which are believed to be BCS-paired in a $^1$S$_0$ state. Then after turning through a sequence of {\em pasta} phases in which the nuclei, driven by the usual competition between surface and Coulomb energies, become rod and then sheet shaped, the crust dissolves into 
a sea of free neutrons, with a smaller fraction of free protons and electrons.   Several features 
of the hadronic liquid interior remain unsolved.   First nuclear three-body forces, which are required within a picture of interacting neutrons and protons to give the correct nuclear saturation density, $n_0$, remain uncertain, particularly with increasing density.  Recent work using chiral perturbation theory, described by Weise \cite{weise}, is helping to determine the three-body forces up to a few times $n_0$.  In addition we do not as yet have calculations of neutron rich nuclear matter as accurate as those for pure neutron matter (pnm) and symmetric nuclear matter (snm -- with equal numbers of neutrons and protons).  Rather one generally relies on an uncertain quadratic interpolation (at all baryon densities, $n_B$) in the proton fraction, $x \equiv n_p/(n_n+n_p)$, of the matter, assuming that the energy density, $E(x)$ depends on $x$ as  $E_{snm} + (2x-1)^2(E_{pnm}-E_{snm})$; here $n_p$ and $n_n$ are the proton and neutron densities.   One consequence of these uncertainties is that the density at which the crust dissolves is determined only to within some 
10\%.

     While one can mathematically construct neutron star equations of state in terms of interacting nucleons up to densities 5-10 $n_0$, e.g., the APR (or AP4) equation of state \cite{APR}, one must ask whether it makes sense physically to take nucleons as the fundamental degrees of freedom at such densities, and build a description of high density matter based on nucleon-nucleon scattering data.     To use a condensed matter analogy, imagine that one had a complete knowledge of niobium-niobium atomic scattering; to what extent could one predict the properties of niobium metal, and especially that it is a superconductor?  The fundamental degrees of freedom at high densities are quarks, and they must be taken seriously in describing the matter in neutron stars!

\subsection{Quarks in neutron stars}  

  A long-standing belief is that quark matter, when it appears in neutron stars, is too soft to support high masses against 
gravitational collapse, and as a consequence quark cores in neutron stars must be small \cite{chin}.  The origin of this idea is the approach of looking for a first order transition between high density hadronic matter and quark matter by seeing which of the two phases has a lower energy density, as a function of baryon density, as illustrated in Fig. 3.   This construction has two problems.  The first is that it assumes that there exists a good hadronic matter equation of state at high baryon densities, despite the problem that hadrons are no longer the correct degrees of freedom at high densities where they substantially overlap.  The second is that the only quark equations of state that one allows in this approach are those that lie under the hadronic equation of state at high density, thus immediately eliminating stiff quark equations of state.    The allowed soft quark equations of state cannot support two solar mass neutron stars.  The consequence is that
typically one finds that quark matter begins to appear only at very high densities, which would barely be reached
even in high mass neutron stars.  Thus this approach allows at  most small quark matter cores, as in, e.g., Ref.~\cite{APR}.

\subsection{The QHC18 equation of state}

    As first pointed out by Sch\"afer and Wilczek \cite{qhcontinuity}, hadronic matter can be connected continuously to quark matter -- the concept of quark-hadron continuity --  in the absence of a change in symmetry between the two phases.   The Monte Carlo calculations of matter at zero baryon chemical potential by the Wuppertal-Budapest \cite{WB} and the HotQCD \cite{Hot QCD} lattice QCD collaborations, which find a continuous crossover with increasing temperature from the hadronic vacuum to the higher temperature quark-dominated vacuum, well illustrate this continuity.  These calculations imply that if there is a Asakawa-Yazaki critical point in the temperature vs.~baryon chemical potential phase diagram (the target of the beam energy scans at RHIC and the SPS), one can circle around the point from the hadronic to the quark side smoothly, as one does around the critical point in water at 373 C from the vapor to the liquid phase.   Hot quark matter and hadronic matter at lower density are continuously connected.

\begin{figure}[h]
\begin{center}
\includegraphics[width=8cm]{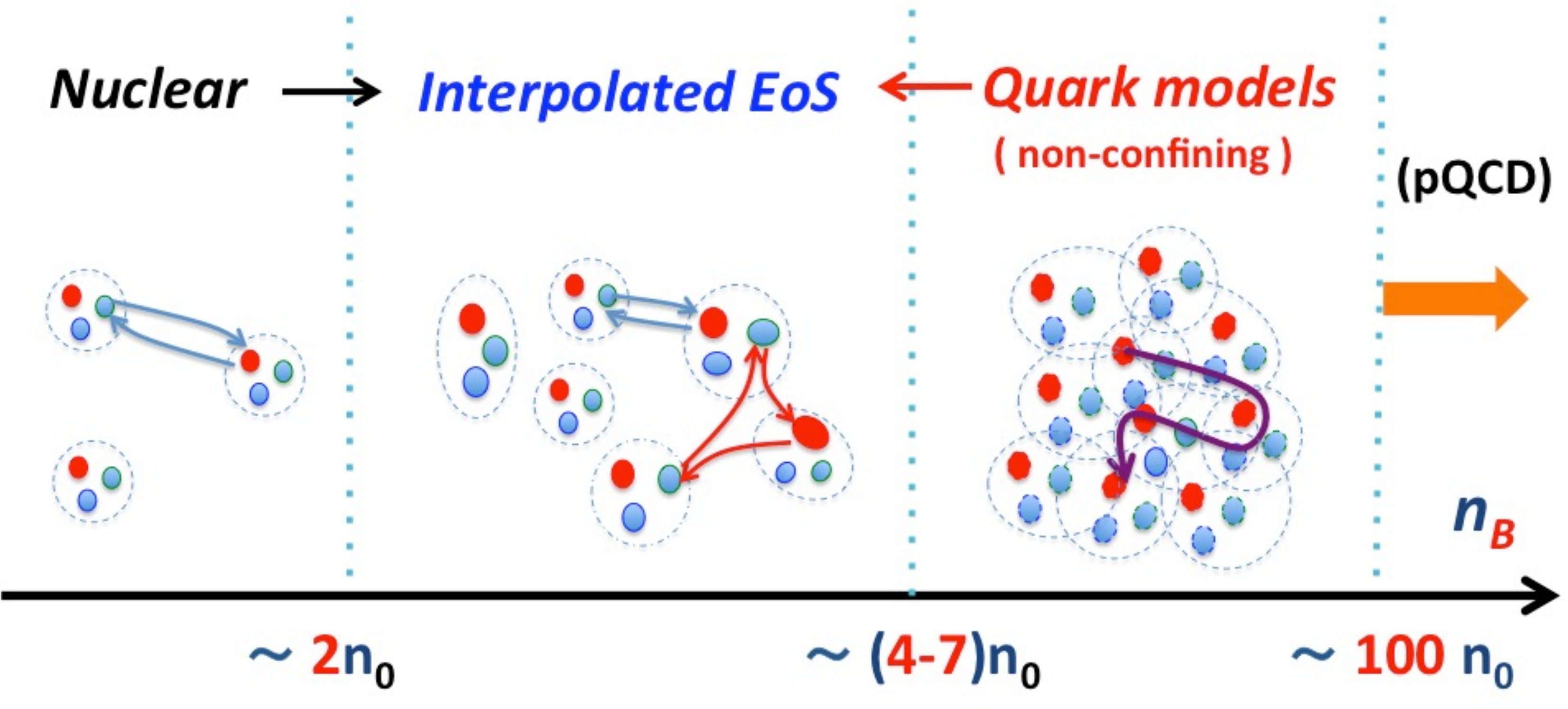}
\end{center}
\vspace{-6pt}
\caption{\footnotesize{Schematic illustration of the transition from nuclear to deconfined quark matter with increasing density.  At densities  $\lesssim 2n_0$, quarks appear in the internucleon interactions, involving relatively few meson or quark exchanges; for $2n_0 \lesssim n_B \lesssim$ 4-7 $n_0$, many-quark exchanges gradually dominate as the system changes from hadronic to quark matter; and for $n_B \gtrsim$ 4-7 $n_0$, quarks no longer belong to specific baryons.  Perturbative QCD is valid only for  at much higher densities.  Figure courtesy of T. Kojo.}}  
\label{fig:3-window}
\end{figure}

        The situation at the very low temperatures one encounters in neutron stars is more complicated owing to the change in symmetry between BCS-paired neutron matter and higher density quark matter paired in a color-flavor locked state.   Jointly, my group and that of Hatsuda \cite{NSReview} have developed an approach based on relatively smooth quark-hadron continuity as the density increases, with at most a weak first order phase transition.   Matter at lower densities is described by a hadronic equation of state, up to perhaps twice $n_0$, and at high densities, above some 4-7 $n_0$,  by a quark matter equation of state;  while the intervening equation of state is uncertain, one can interpolate between these two limiting regimes.  The high density equation of state and its interpolation to the hadronic regime are in fact highly constrained by two requirements, that the resulting equation of state be sufficiently stiff to support two solar mass neutron stars, and that the matter be stable against density fluctuations.  
                
      We determine the equation of state in pressure, $P$, vs. baryon chemical potential, $\mu_B$.  The advantage of this choice of thermodynamic variables is that $P$ must be a continuous non-decreasing function of $\mu_B$.   A first order phase transition appears as a change in slope, $\partial P/\partial\mu_B = n_B$, the baryon density.   Specifically we have used the APR hadronic equation of state, and more recently the Togashi equation of state \cite{hajime} up to 2 $n_0$ (which contains the same physics as APR, now including in the inner crust), and the  Nambu--Jona-Lasinio (NJL) model \cite{NSReview} to describe three-flavor quark matter, thus including strangeness in the mix.   Between these two limits we do a simple polynomial interpolation in $P$ vs. $\mu_B$ .  
      
       The  Lagrangian density of the NJL model is ${\cal L} = 
 \overline{q} ( \gamma^\mu p_\mu  - \hat{m}_q +\mu_q \gamma^0) q + {\cal L}^{(4)} + {\cal L}^{(6)}$, where $q$ is the quark operator, and $\hat{m}_q$ is the bare quark mass matrix; ${\cal L}^{(4)}$ includes the usual four quark interaction with coupling $G_s$ (of order 3-5 GeV$^{-2}$), plus two phenomenological interactions, a short range repulsive universal vector interaction, with Lagrangian ${\cal L}_V =  -g_V( \bar q  \gamma^\mu q)^2$ between quarks, and a BCS or diquark pairing interaction, schematically, ${\cal L}_H \sim H d^\dagger d$, where $d$ is the diquark pair operator; and finally $ {\cal L}^{(6)}$ describes the six-quark Kobayashi-Miskawa-'t Hooft  interaction representing the effects of the
instanton-induced QCD axial anomaly which breaks the U(1)$_A$ axial symmetry of the QCD Lagrangian.  Full details are in Ref.~\cite{NSReview}.  In the quark matter equation of state we treat $g_V$ and $H$ as phenomenological parameters.  

   The name, QHC,  of the equation of state indicates the crossover from hadrons to quarks, with QHC18($g_V,H$) indicating the latest iterations in the past year.  The full QHC18 equation of state is detailed in Ref.~\cite{NSReview} and is given, together with instructions for use, on the websites: Home Page of 
Relativistic EOS table for supernovae \url{http://user.numazu-ct.ac.jp/~sumi/eos/index.html#QHC18}  and the CompOSE repository \url{https://compose.obspm.fr/home/}.  
\begin{figure}[h]
\begin{center}
\includegraphics[width=9cm]{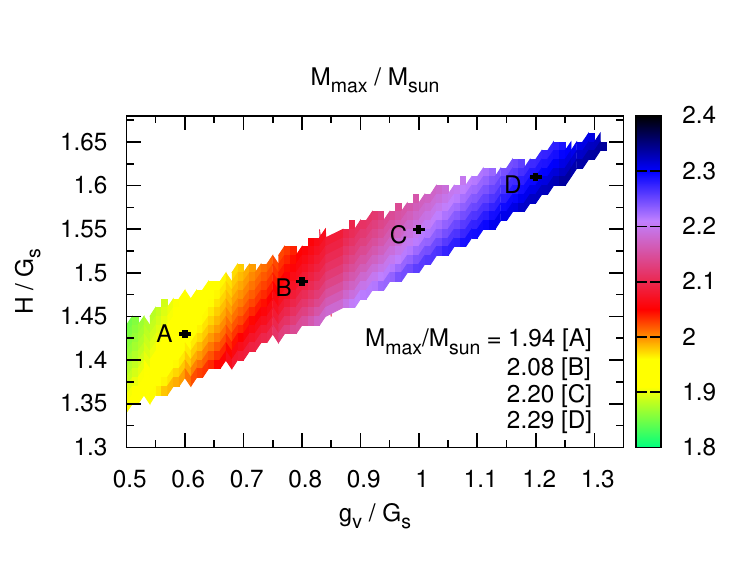}
\end{center}
\vspace{-18pt}
\caption{\footnotesize{The regions in parameter space $\{g_V,H\}$ (in units of the NJL coupling constant $G_s$) that yield consistent equations of state. The color scale indicates the maximum neutron star mass supported by the corresponding 
QHC($g_V,H$) equation of state.   Figure courtesy of T. Kojo.}}  
\label{gVH}
\end{figure}  

    The minimal model, $g_V=H=0$, yields an equation of state which is too soft to support 2$M_\odot$ neutron stars. Increasing the repulsion alone stiffens the equation of state; however at the same time it develops an inflection point in $P$ vs. $\mu_B$, indicating a mechanical instability, $\partial^2 P/\partial \mu_B^2 =   \partial n_B/\partial \mu_B <0$.   Only by simultaneously including a sufficiently large diquark pairing $H$, can one construct an equation of state satisfying the constraints.    Figure~1 shows the equation of state in $P$ vs. $n_B$, for different parameter sets $\{g_V,H\}$ in Fig.~5.    The QHC18 equation of state for the parameter range from C to D agree quite remarkably well with the equation of state in $P$ vs. $n_B$ inferred by LIGO and Virgo \cite{Abbott2018}.   As Fig.~5 shows, the region in the parameter set $\{g_V, H\}$ plane that gives stable neutron stars is narrow.  
        
\begin{figure}[h]
\begin{center}
\includegraphics[width=8cm]{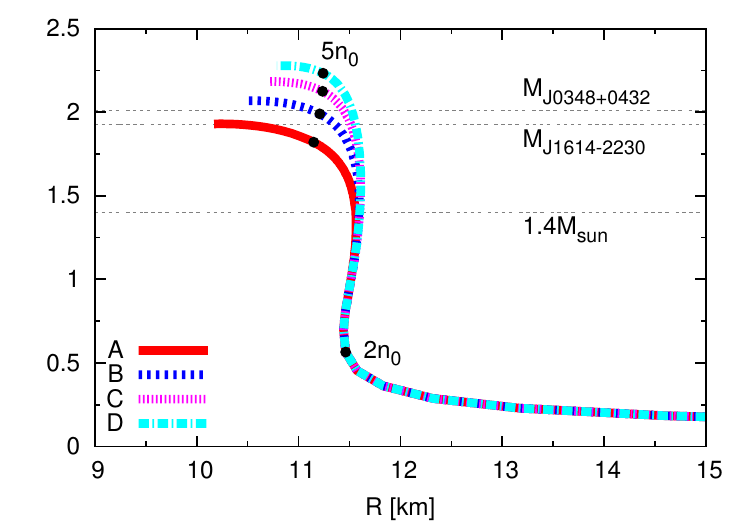}
\end{center}
\caption{\footnotesize{Mass vs. radius for the equation of state QHC18.  The parameter sets A through D are indicated in the inset in Fig. 1.  The black dots show the stars where the central density is 5$n_0$.   Figure courtesy of T. Kojo.}}  
\label{MvsR}
\end{figure}

    The range of pairing parameters $H$, above 1.5 $G_s$ are larger than conventionally included 
value 3/4 $G_s$ in the NJL model.   Such larger phenomenological pairing strengths are not surprising,
since with decreasing density two and three quark correlations build up in quark matter,
eventually becoming well-defined nucleons at lower densities.  
         
    Figure 6 shows the run of neutron star masses vs. stellar radius calculated with QHC18 for parameter sets A through D.   The maximum neutron star mass for parameter set A is only 1.94 $M_\odot$, while B, C, and D can support 2 $M_\odot$.   The radii are around 11.5 km.

\section{Conclusion} 
 
   The study of neutron stars is rapidly becoming an important subfield of high energy nuclear physics, an outcome of the observational consequences of the microscopic physics of dense matter.   While the QHC18
equations of state do treat quarks at high densities, there remain significant uncertainties, including those 
in the hadronic equation of state, and in the interpolation between hadronic matter and quark matter, which affect the maximum neutron star masses and radii.   Importantly, we lack a theory of the crossover from hadronic to quark matter.  The intermediate regime may be amenable to a description of gauge invariant mesonic and baryonic-like excitations in quark matter \cite{yifan} which complements the picture presented by Weise \cite{weise} of matter at densities above 2 $n_0$ in terms of hadronic excitations.   We need to understand microscopically the vector coupling, $g_V$, and pairing strengths $H$ used in QHC18 including their density dependences (Ref.~\cite{exchange} argues that one can understand the effective density-dependent vector repulsion as arising from the quark exchange energy at moderate densities, including effects of non-perturbative chiral and diquark condensates in the vacuum, and gluon effective masses).  More generally we would
certainly like to go beyond the phenomenological NJL model, which does not contain explicit gluons.  
  A further challenge is to extend the present QHC equations of state to non-zero temperatures ($\lesssim$ 100 MeV)
to be used in modelling neutron star--neutron star and neutron star--black hole mergers as sources of gravitational 
radiation; work in progress is reported in Ref.~\cite{hotqhc1}.      \\
   
\section*{Acknowledgments} 

   This paper is dedicated to the memory of my old friend and colleague T. Takatsuka, who for half a century was a major player in neutron star physics; he will live on through his impact on the field.    I am indebted as well to my colleagues C. J. Pethick, T. Hatsuda and T. Kojo, and students P. D. Powell and Y. Song, for their invaluable contributions to the understanding of neutron star physics discussed in this paper.  The research described here was supported in part by U.S. National Science Foundation Grants PHY1305891 and PHY1714042.

\end{document}